%% file: bot_pap.tex
\documentclass[aps,prd,12pt,preprint,tightenlines,nofootinbib]{revtex4-1}

\usepackage[caption=false]{subfig}
\usepackage{epsfig}
\usepackage{hyperref}
\usepackage{amssymb}
\usepackage{color}
\usepackage{slashed}
\usepackage{amsmath}
\newcommand{\be}{\begin{equation}}
\newcommand{\ee}{\end{equation}}
\newcommand{\ba}{\begin{eqnarray}}
\newcommand{\ea}{\end{eqnarray}}
\newcommand{\bd}{\begin{displaymath}}
\newcommand{\ed}{\end{displaymath}}

\newcommand{\commentout}[1]{{}}

\begin{document}

\title{Sequential bottomonium production at high temperatures}
\author{Peter Petreczky}
\email{petreczk@quark.phy.bnl.gov}
\affiliation{Physics Department, Brookhaven National Laboratory, Upton, NY 11973, USA}
\author{Clint Young}
\email{youngc@nscl.msu.edu}
\affiliation{Department of Physics and Astronomy and National Superconducting Cyclotron Laboratory, Michigan State University, East Lansing, MI 48824, USA}
\date{\today}

\begin{abstract}
Bottomonium production in heavy ion collisions is modified compared with any simple extrapolation from elementary collisions. This modification is most likely caused 
by the presence of a deconfined system of quarks and gluons for times of several fm/c. In such a medium, bottomonium can be destroyed, but the constituent bottom quarks
will likely stay spatially correlated due to small mean free paths in this system. With these facts in mind, we describe bottomonium formation with a coupled set of equations.
A rate equation describes the destruction of $\Upsilon(1S)$ particles, while a Langevin equation describes 
how the bottom quarks stay correlated for a sufficiently long time so 
that recombination into bottomonia is possible. We show that 
within this approach it is possible to understand the magnitude of $\Upsilon(1S)$ suppression in heavy ion collisions and the larger
suppression of the $\Upsilon(2S)$ state, implying that the reduction in the ratio of $\Upsilon(1S)/\Upsilon(2S)$ yield in heavy
ion collision does not necessarily correspond to sequential melting picture.
\end{abstract}

\maketitle

\section{Introduction}

Quarkonium has long been used to examine the properties of heavy ion collisions (see
Refs. \cite{Mocsy:2013syh,Andronic:2015wma} for recent reviews). 
Within the context of QCD at finite temperature, the most common description of 
the dynamics of quarkonium in these collisions is one where they almost immediately melt: 
the potential between the heavy quarks becomes screened by the deconfined quarks 
and gluons, and the heavy quarks separate from each other \cite{Matsui:1986dk}. 
Some of the earliest measurements at SPS were first interpreted as evidence for ``$J/\psi$ suppression'' 
\cite{Abreu:2000ni}; 
this is now mostly explained by the modifications of the parton distribution functions in colliding nuclei and the absorption cross-section for $J/\psi$ 
collectively called cold nuclear matter effects
\cite{Arnaldi:2007zz}. 
In heavy ion collisions at the Relativistic Heavy Ion Collider (RHIC) and the Large Hadron Collider (LHC), the nuclear modification factor $R_{AA}$ for 
quarkonium production cannot be explained by cold nuclear matter effects \cite{Andronic:2015wma}. 
However, in the case of $J/\psi$ the modification is not 
consistent with instantaneous melting of charmonium states above various temperature thresholds. Another important observation at LHC is
the large suppression of $\Upsilon(2S)$ and $\Upsilon(3S)$ states compared with the $\Upsilon(1S)$ state in heavy ion collisions \cite{Chatrchyan:2011pe}.
This is often interpreted as a signature of sequantial bottomonium melting.

The mistake in the earliest theoretical work could very well be the simple hypothesis for the dynamics of quarkonium, where the heavy quarks, once screened, simply fly apart.
This hypothesis is not supported by any direct evidence: the finite temperature lattice QCD calculations measured the Polyakov loop, which is related to the free energy 
of infinitely heavy fundamental charges (see Refs. \cite{Bazavov:2016qod,Borsanyi:2015yka} for recent calculations) but was not in any way a simulation of quarkonium. 
Another hypothesis for the behavior of quarkonium at finite temperature is inspired 
by the observation that charm mesons have significant elliptic flow at RHIC \cite{Adler:2005ab}, suggesting 
a strong interaction with the surrounding medium. The dynamics of single charm quarks is a diffusive process that can be described by a  
relativistic generalization of the Langevin equation \cite{Moore:2004tg,vanHees:2004gq,vanHees:2005wb}. This model explains the elliptic 
flow of charm mesons when the diffusion coefficient for heavy quarks is sufficiently  small. 
Shortly after the original proposal of a large suppression of $J/\psi$ particles in heavy ion collisions, it was noticed that 
on the contrary, the drag which charm quarks experience might lead to an enhancement instead \cite{Svetitsky:1987gq}.
A simple but powerful model for quarkonium production in heavy ion collisions 
can be made from the Langevin equation, where a heavy quark and 
anti-quark interact with each other according to a screened Cornell potential and interact, independently, with the surrounding medium, experiencing both drag and rapidly 
decorrelating random forces \cite{Young:2008he}. This model was first shown to describe the existing data on $J/\psi$ production at RHIC \cite{Young:2008he,Young:2009tj}.
and then made successful predictions of the results at the LHC \cite{Young:2011ug}. One of the 
greatest strengths of this model is its simplicity in implementation in event generators for heavy ion collisions \cite{Young:2011ug}: with it, 
heavy quarks as well as quarkonium and even 
exotic $B_c$ mesons are described simultaneously. 

However, 
this model clearly has limitations. One serious limitation is the independence of the interaction of the heavy quark with the medium and the interaction of the anti-quark 
in the same pair with the medium. 
One would expect there to be significant correlation between the interactions, 
especially when the pair is tightly bound and the spatial separation between the quarks is very small. 
This is the situation for bottomonium in the ground state which has a typical size that is smaller than the size of the $J/\psi$ and
excited bottomonium states.  Therefore, the tightly bound $\Upsilon(1S)$ should be treated differently.

The time scale of bound state formation is also very important.
It is usually assumed that the formation of quarkonium states happens before the formation 
of the deconfined medium. The time to form the bound state is typically larger than the inverse of the binding
energy, $t_{form} > 1/(m v^2)$, where $v$ is the heavy quark velocity inside the bound states.
For ground state bottomonium $v^2 \simeq 0.1$ and for bottom quark mass $m_b \simeq 5 $ GeV we get
$t_{form} > 0.5$ fm. This is not too different from the time scale of formation  of the deconfined thermalized
medium assumed in hydrodynamics simulations. So there is no clear separation between bound state formation and
the formation time scale of the hot medium. Since $\Upsilon(1S)$ will have a thermal width in the deconfined medium
it will be dissociated. The $b$ and $\bar b$ emerging from
the dissociation will remain correlated, however. 
This correlated pair is described by Langevin dynamics.
If the correlation between the $b$ quark and anti-quark persists, there is a possibility that
the $b \bar b$ pair will form 
$\Upsilon(1S)$ as well as the excited bottomonium states at later stages, when
the system cools down.

In this paper we apply the above idea to study bottomonium production in heavy ion collisions. We couple
the rate equation to the Langevin dynamics of $b\bar b$ pair in the hot medium and study their time evolution.
The coupled equations might explain the observed suppression pattern in a way similar to sequential melting:
the quarks in the excited states can diffuse farther away from each other than the quarks initially in a 
tightly bound ground state, described initially with the rate equation.

\section{The coupled rate and Langevin equations for bottomonium}
\label{coupled}

We assume that all $b\bar b$ pairs are produced in the initial hard
collisions.  We also assume that the system produced in heavy ion collisions
is rapidly thermalized and the produced heavy quark pairs immediately 
undergo multiple scatters in the medium. The hard production of the $b\bar b$
and their initial spectrum is calculated using PYTHIA 8.1 \cite{Sjostrand:2007gs}.
We use the default value $M_b=4.5$ GeV for the bottom quark mass in the simulations.
It is well known that only a small fraction of the produced heavy quark pairs makes
quarkonium. In our exploratory study we separate the initially 
produced $b\bar b$ pairs in three different bins according to their center of mass
energy: $E <9.3 {\rm GeV}$,~$9.3 {\rm GeV}< E < 9.5 {\rm GeV}$, and $E>9.5$ GeV.
This binning is similar to the idea of color evaporation model 
for quarkonium production \cite{Fritzsch:1978kn,Amundson:1996qr}.
We assume that $b\bar b$ pairs in the first energy bin form $\Upsilon$ instantly, while 
the pairs in the third bin correspond to the open beauty sector. The pairs in the second bin
will be treated as correlated pairs described by Langevin dynamics. We will study the dynamics
of these pairs in the hot medium as function of time and see up to which time scales the correlation
persists. Once the system produced in heavy ion collisions cools down sufficiently and hadronizes,
the remaining correlated pairs will form different bottomonium bound states.

To describe the dynamics of the correlated 
$b \bar b$ pairs we use the Langevin dynamics of the $b$ ($\bar b$) quark inside
the $b \bar b$ pair defined by Langevin equation
\begin{equation}
\dot{p}^i = -\eta p^i + \xi^i(t) - \nabla^i V({\bf x}).
\end{equation}
Here $\xi(t)$ is the random force from the medium acting on the $b$ quark and $V({\bf x})$ is
the potential between the b quark and the anti-quark.
The random force satisfies the condition $\langle \xi^i(t) \xi^j(t')\rangle=\kappa \delta^{ij} \delta(t-t')$.
The coefficient $\kappa$ and the drag coefficient $\eta$ are related to each other and to the 
diffusion constant $D$ in coordinate space:
\begin{equation}
\eta=\frac{\kappa}{2 M T},~~D=\frac{T}{M \eta}=\frac{2 T^2}{\kappa}.
\end{equation}
Roughly speaking the diffusion constant $D$ corresponds to the mean free path
of light degrees of freedom in the deconfined medium. In the weak coupling limit
it scales like $D \sim 1/(g^4 T)$ and thus could be quite large {\cite{Moore:2004tg}.
The mean free path of the heavy quark is always larger than for the light degrees
of freedom and scales like $D \cdot M/T$. In the strong coupling limit the 
diffusion constant $D$ is small, $D \cdot 2 \pi T \simeq 1$ \cite{CasalderreySolana:2006rq}.
We assume that the formed deconfined medium is strongly coupled and choose 
$2\pi TD = 1.5$. The use of Langevin dynamics for the quark anti-quark pair can be justified
if the binding energy is small \cite{Akamatsu:2014qsa}.

We need to specify the quark anti-quark potential. At zero temperature it is
well known and can be parameterized by the Cornell form. We use the parameterization
of the Cornell potential based on lattice QCD calculations \cite{Petrov:2006pf}.
The potential at non-zero temperature is not known well, although there are ongoing
attempts to calculate it on the lattice \cite{Bazavov:2014kva,Burnier:2014ssa}.
Therefore, for $V$ we choose the so-called maximally binding potential \cite{Mocsy:2007jz}.
It is constructed in the following way. At distances $r<r_{scr}(T)$ it coincides
with the $T=0$ Cornell form, while for $r>r_{scr}(T)$ it is simply equal to constant
$V_{\infty}(T)$. Here $r_{scr}(T)$ is the screening radius which is chosen to be $0.8/T$
because the lattice QCD calculations show that the singlet free energy is exponentially screened
for $r>0.8/T$ \cite{Bazavov:2012fk}. Requiring that the potential is continuous at $r=r_{scr}(T)$
we fix the value of $V_{\infty}(T)$. This completely specifies the potential.
\begin{figure}
\includegraphics[width=14cm]{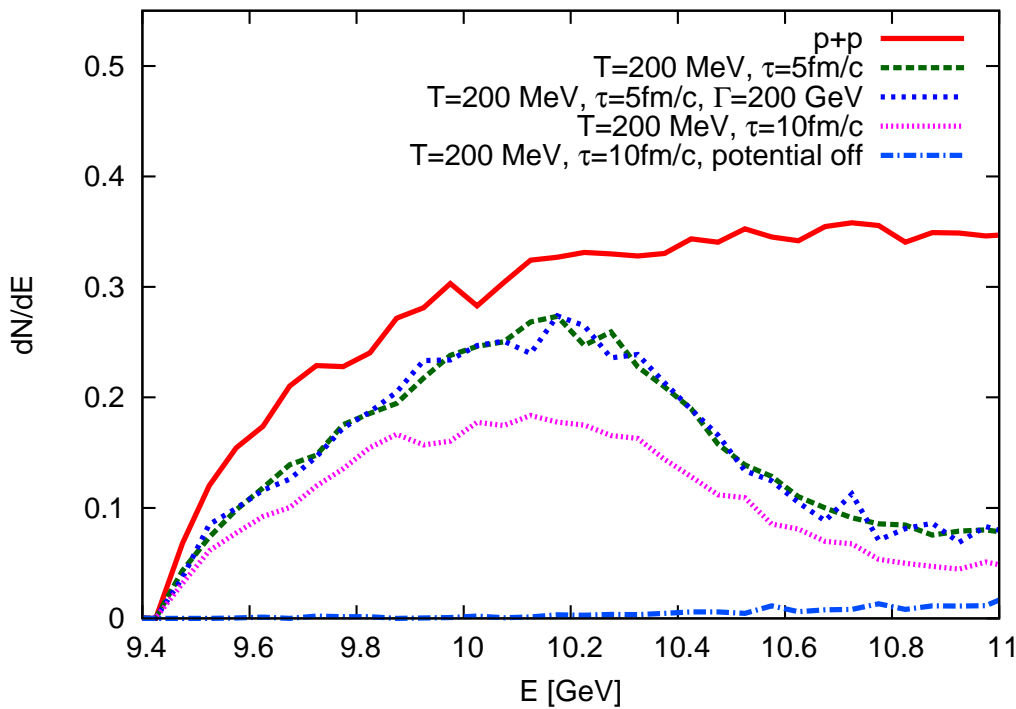}
\caption{The distribution of $b \bar b$ pairs in pp collisions and in
heavy ion collisions in center of mass energy. It is assumed that the medium
has a fixed temperature (here $T=200$ MeV). Results for different 
life times of the deconfined medium are shown as well for scenarios when the feed-down
is included or the quark anti-quark potential is turned off.}
\label{fig:dNdE}
\end{figure}

Since the $\Upsilon(1S)$ state is tightly bound it may exist as a bound state even
at the highest temperatures that can be achieved in heavy ion collisions and form early.
Therefore, we will treat the $1S$ bottomonium as a distinct particle that exists in the
deconfined medium. Since the $1S$ bottomonium has a thermal width it will be dissociated
in the medium and the number of these particles at each point of time is determined by
rate equation
\begin{equation}
\frac{d N_{\Upsilon}}{d t}=-\Gamma(T)  N_{\Upsilon} {\rm .}
\end{equation}
The thermal width $\Gamma(T)$ was calculated in the weak coupling approach for bound states of
infinitely heavy quarks \cite{Brambilla:2008cx} and for realistic values 
of the strong coupling constant corresponds $\Gamma(T) \sim 0.1 T$. Calculations
have been extended to finite quark mass 
\cite{Brambilla:2010vq,Brambilla:2011sg,Brambilla:2013dpa}; one gets similar numerical values for $\Gamma$.
Since the highest temperature considered in our study is $T=400$ MeV we will use $\Gamma=40$ MeV for
the thermal width. Since the validity of weak coupling calculations is not clear, especially 
close to $T_c$ in addition we also performed calculations with a larger width of $\Gamma=200$ MeV. 
This value of the width, which is typical hadronic width was used also at lower temperature of $T=200$ MeV.
Once the $1S$ bottomonium is dissolved, the resulting $b \bar b$ pairs is counted as a correlated pair
since the relative energy of the quark and anti-quark is small.
Thus, we have feed-down from bottomonium sector to the sector of correlated $b\bar b$ pairs.

With this model we study the bottomonium formation in deconfined medium assuming that it
has a constant temperature. First, we study how the distribution of $b \bar b$ pairs
in the energy is affected by the medium. Our findings are shown in Fig. \ref{fig:dNdE}.
We see that the distribution increases monotonically with increasing energy in the studied energy range
in the proton-proton (pp) collisions, and eventually reaching a maximum. The presence of the medium modifies the initial
distribution significantly: the distribution has a clear peak around $E=10$ GeV already at $t=5fm/c$. 
The distribution changes very slowly in time,  the shape of the distribution remains the same around the peak, only
the high energy tail changes. This appears as a change of normalization of the distribution around the peak.
We also included the feed-down from $1S$ bottomonium in the calculations but this does not change the distribution
in any visible way (c.f. Fig. \ref{fig:dNdE}). The presence of the potential between the quark and anti-quark
on the other hand has a huge effect on the distribution as can be seen from Fig. \ref{fig:dNdE}. In absence of the potential 
the peak of the distribution shifts to much larger energies.

\begin{figure}[ht]
  \centering
  \includegraphics[width=13cm]{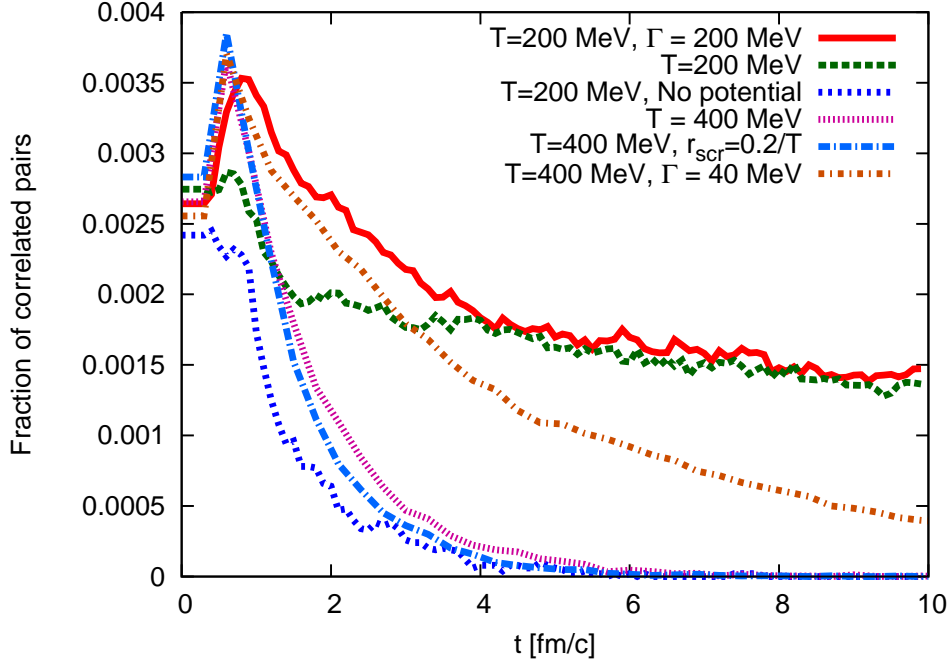}
  \caption{The fraction of correlated $b\bar b$ pairs
  as function of life-time of the deconfined medium. Different lines
  show the effect of feed-down from $1S$ bottomonium states for different thermal width and the presence
  of or the shape of the quark anti-quark potential.
  Results are shown for two different temperatures of the deconfined medium, $T=200$ MeV and $T=400$ MeV.
} 
\label{fig:frac}
\end{figure}

We could examine the evolution of correlated $b\bar b$ pairs, defined as pairs with energy $9.3{\rm GeV} < E < 9.5 {\rm GeV}$
with the time. This is shown in Fig. \ref{fig:frac}. An interesting feature seen in the figure is the spike at small times.
This spike is due to the fact that the dissociation of the $1S$ bottomonium states leads to an increase in the correlated
$b\bar b$ pairs. The evolution of the correlated $b \bar b$ pairs strongly depends on the presence and shape of the potential.
We consider evolution scenarios, where the potential is turned off or is very short range, namely we assume that $r_{scr}=0.2/T$.
In both cases the fraction of correlated pairs decreases rapidly with time. 
If the potential is present the fraction of correlated $b\bar b$ pairs decreases very slowly with time. For example
if the deconfined medium lives for 10 fm the fraction of correlated pairs is reduced only by factor of two assuming a temperature
$T=200$ MeV.
While at early times the coupling to rate equation
has a significant effect at later time this coupling has essentially no effect. 
This can be seen in Fig. \ref{fig:frac} for the case $T=200$ MeV. 
In other words, the observed bottomonium yields
in heavy ion collisions will not depend on the assumption of whether the $1S$ state is formed early or after the medium is thermalized.
\begin{figure}[ht]
  \centering
  \includegraphics[width=13cm]{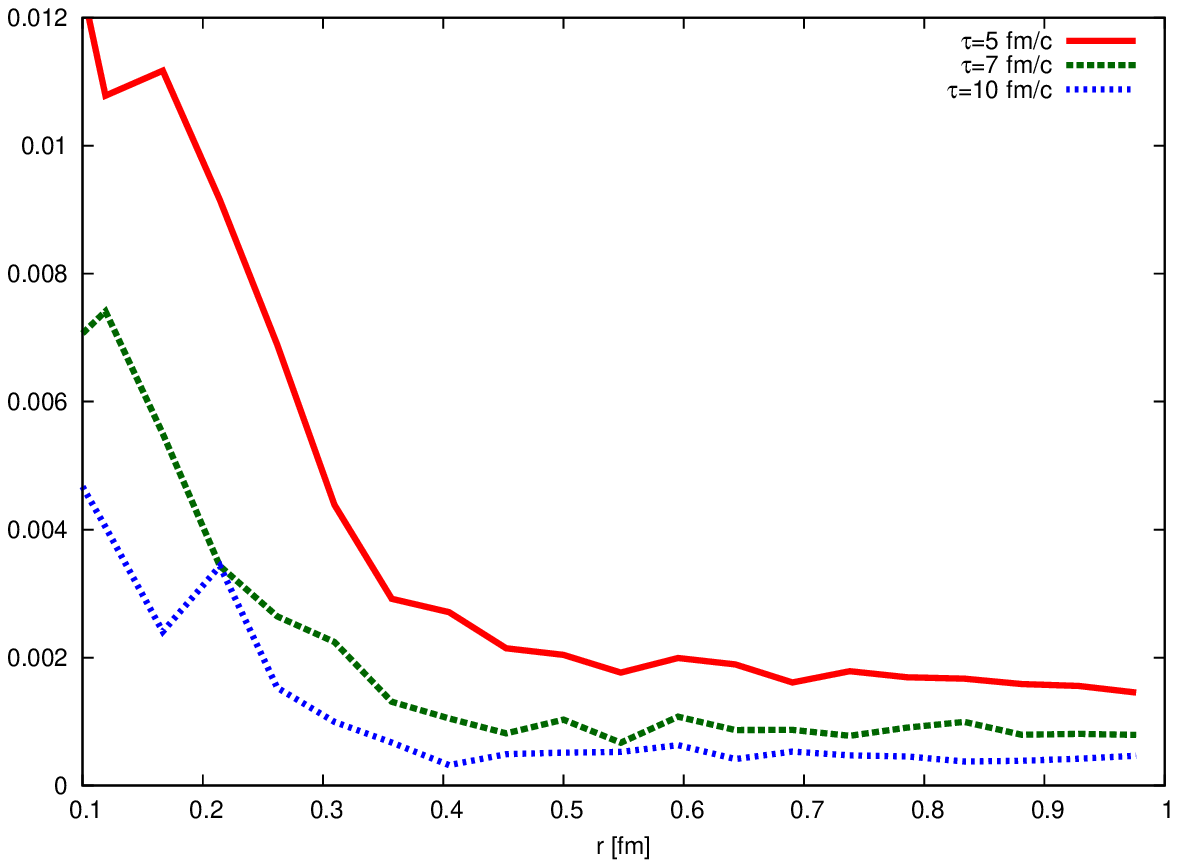}
  \caption{The distribution of the correlated $b\bar b$ 
   pairs as function of the relative distance between the quark and
   and anti-quark for a deconfined medium at $T=200$ MeV. The wiggles in the figure are due to limited statistics.}
  \label{fig:rT400}
\end{figure}

It is also useful to study the distribution of the correlated $b \bar b$ pairs as a function of the relative distance
between the quarks,
especially when one wants to study bottomonium bound state formation from the correlated $b \bar b$ pairs. This distribution 
will determine the probability that a given bound state is formed from the correlated pair. The distribution of $b \bar b$
pairs as a function of the relative distance is shown in Fig. \ref{fig:rT400} for different times. While the distribution
changes with increasing time its shape is almost unaffected. It is the overall magnitude of the distribution that is decreasing,
reflecting the fact the total number of correlated $b\bar b$ pairs is decreasing with time, while the number of uncorrelated pairs is
increasing. 
This feature was first noticed for charm quarks \cite{Young:2008he}.
The most prominent feature of the distribution is that it has a peak at small distances and the width of this peak
is about $0.25$ fm. There is also a long tail for large relative distances. The shape of the distribution implies that 
if the system would cool  down rapidly to a temperature where various bottomonium states could form the correlated  
$b \bar b$ pairs would mostly form $1S$ state rather than excited bottomonium states since the size of $1S$ state is about $0.25$ fm.
We see a sequntial formation pattern: smaller bottomonium ground state is more likely to be formed than larger excited
bottomonium states.
Therefore, the stronger suppression of $\Upsilon(2S)$ relative to $\Upsilon(1S)$ in heavy ion collisions observed by
CMS is not necessarily related to sequential melting of bottomonium states.

\section{Conclusions}
In this paper we considered bottomonium formation in hot deconfined medium that may be produced in relativistic heavy
ion collisions using Langevin dynamics for correlated $b \bar b$ pairs coupled to rate equation for $1S$ bottomonium.
We showed that a large fraction of quark anti-quark pairs that were correlated during the initial hard production will
remain correlated in the hot medium for rather long time of about $5-10$ fm/c. This is the typical life-time of the system
created in heavy ion collisions. 
We studied in detail the sensitivity of these correlation to the form of the quark anti-quark potential.
The distribution of the correlated $b \bar b$ pair in relative distance is such, that
it will dominantly form $1S$ bottomonium. This may explain the suppression of the relative yield of $\Upsilon(2S)/\Upsilon(1S)$
observed by CMS without invoking the idea of sequential suppression; one could say that there is a sequential pattern of
bottomonium formation according to their size.

Our discussion was quite simplified as we studied
production in a static medium. It is relatively straightforward to extend this approach to expanding realistic medium as it was
done in the case of charmonium \cite{Young:2011ug}. 
We plan to do so in the near future using the same framework as in Ref. \cite{Young:2011ug}.
However, there are many theoretical problems that need to be resolved.
The Langevin dynamics is purely classical and ignores quantum effects. As long as the binding energy is very small
neglecting quantum effects may not be too bad of an approximation. 
But as the temperature decreases binding force between the quark anti-quark
becomes stronger and quantum effects will be essential. There have been several suggestions on how to incorporate
quantum effects in the dynamics of correlated heavy quark anti-quark pair 
\cite{Young:2010jq, Buyukdag:2015sba, Akamatsu:2011se,Rothkopf:2013ria,Rothkopf:2013kya,Blaizot:2015hya,Gossiaux:2016htk}
Furthermore, close to the transition temperature non-perturbative effects
will become important, so the effective field theory approach needs to be generalized to strong coupling and 
the potential will have to be calculated on the lattice. In the vicinity of the transition temperature various
bound states, including excited bottomonium states \cite{Kim:2014iga} and open heavy flavor hadrons \cite{Mukherjee:2015mxc}
may exist, which will influence the chemistry of bottomonium production.  In this region the simple
Langevin dynamics will not be sufficient and a coupled set of equations will have to be considered.
However, even at early stages of the dynamics refinements will be necessary. We assumed a fixed mass for $b$ quark
in PHYTIA calculations as well as in the Langevin dynamics. This mass should be matched to the constituent
bottom quark mass or the pole mass and eventually to the in-medium temperature dependent bottom quark mass.
This may effect the energy distribution of the produced $b \bar b$ pair significantly.
Such tuning of the bottom quark mass will be essential to define energy bins that give correct yields
in pp collisions and correspond to the fractions of $b \bar b$ pairs that produce bottomonium in
color evaporation model. 

It is clear that the problem of bottomonium production in heavy ion collisions is highly non-trivial and requires lots of further work
before meaningful comparison between theory and experiment could be done.

\section*{Acknowledgement}
This work was supported by the Department of Energy 
through grant number DE-FG02-03ER41259 and 
Contract No. DE-SC0012704.

\input{disclaimer}

\bibliography{ref}

\end{document}

%% file: disclaimer.tex
\noindent
{\bf DISCLAIMER}
\noindent
This report was prepared as an account of work sponsored by an agency of the United States
Government. Neither the United States Government nor any agency thereof, nor any of their
employees, nor any of their contractors, subcontractors, or their employees, makes any warranty,
express or implied, or assumes any legal liability or responsibility for the accuracy, completeness,
or any third party's use or the results of such use of any information, apparatus, product, or
process disclosed, or represents that its use would not infringe privately owned rights. Reference
herein to any specific commercial product, process, or service by trade name, trademark,
manufacturer, or otherwise, does not necessarily constitute or imply its endorsement,
recommendation, or favoring by the United States Government or any agency thereof or its
contractors or subcontractors. The views and opinions of authors expressed herein do not
necessarily state or reflect those of the United States Government or any agency thereof.

%% file: bot_pap.bbl
\begin{thebibliography}{39}%
\makeatletter
\providecommand \@ifxundefined [1]{%
 \@ifx{#1\undefined}
}%
\providecommand \@ifnum [1]{%
 \ifnum #1\expandafter \@firstoftwo
 \else \expandafter \@secondoftwo
 \fi
}%
\providecommand \@ifx [1]{%
 \ifx #1\expandafter \@firstoftwo
 \else \expandafter \@secondoftwo
 \fi
}%
\providecommand \natexlab [1]{#1}%
\providecommand \enquote  [1]{``#1''}%
\providecommand \bibnamefont  [1]{#1}%
\providecommand \bibfnamefont [1]{#1}%
\providecommand \citenamefont [1]{#1}%
\providecommand \href@noop [0]{\@secondoftwo}%
\providecommand \href [0]{\begingroup \@sanitize@url \@href}%
\providecommand \@href[1]{\@@startlink{#1}\@@href}%
\providecommand \@@href[1]{\endgroup#1\@@endlink}%
\providecommand \@sanitize@url [0]{\catcode `\\12\catcode `\$12\catcode
  `\&12\catcode `\#12\catcode `\^12\catcode `\_12\catcode `\%12\relax}%
\providecommand \@@startlink[1]{}%
\providecommand \@@endlink[0]{}%
\providecommand \url  [0]{\begingroup\@sanitize@url \@url }%
\providecommand \@url [1]{\endgroup\@href {#1}{\urlprefix }}%
\providecommand \urlprefix  [0]{URL }%
\providecommand \Eprint [0]{\href }%
\providecommand \doibase [0]{http://dx.doi.org/}%
\providecommand \selectlanguage [0]{\@gobble}%
\providecommand \bibinfo  [0]{\@secondoftwo}%
\providecommand \bibfield  [0]{\@secondoftwo}%
\providecommand \translation [1]{[#1]}%
\providecommand \BibitemOpen [0]{}%
\providecommand \bibitemStop [0]{}%
\providecommand \bibitemNoStop [0]{.\EOS\space}%
\providecommand \EOS [0]{\spacefactor3000\relax}%
\providecommand \BibitemShut  [1]{\csname bibitem#1\endcsname}%
\let\auto@bib@innerbib\@empty
%</preamble>
\bibitem [{\citenamefont {Mocsy}\ \emph {et~al.}(2013)\citenamefont {Mocsy},
  \citenamefont {Petreczky},\ and\ \citenamefont {Strickland}}]{Mocsy:2013syh}%
  \BibitemOpen
  \bibfield  {author} {\bibinfo {author} {\bibfnamefont {A.}~\bibnamefont
  {Mocsy}}, \bibinfo {author} {\bibfnamefont {P.}~\bibnamefont {Petreczky}}, \
  and\ \bibinfo {author} {\bibfnamefont {M.}~\bibnamefont {Strickland}},\
  }\bibfield  {journal} {\bibinfo  {journal} {Int. J. Mod. Phys.}\ }\textbf
  {\bibinfo {volume} {A28}},\ \bibinfo {pages} {1340012} (\bibinfo {year}
  {2013}),\ 1302.2180\BibitemShut {NoStop}%
%%CITATION = ARXIV:1302.2180;%%
\bibitem [{\citenamefont {Andronic}\ \emph {et~al.}(2016)\citenamefont
  {Andronic} \emph {et~al.}}]{Andronic:2015wma}%
  \BibitemOpen
  \bibfield  {author} {\bibinfo {author} {\bibfnamefont {A.}~\bibnamefont
  {Andronic}} \emph {et~al.},\ }\bibfield  {journal} {\bibinfo  {journal} {Eur.
  Phys. J.}\ }\textbf {\bibinfo {volume} {C76}},\ \bibinfo {pages} {107}
  (\bibinfo {year} {2016}),\ 1506.03981\BibitemShut {NoStop}%
%%CITATION = ARXIV:1506.03981;%%
\bibitem [{\citenamefont {Matsui}\ and\ \citenamefont
  {Satz}(1986)}]{Matsui:1986dk}%
  \BibitemOpen
  \bibfield  {author} {\bibinfo {author} {\bibfnamefont {T.}~\bibnamefont
  {Matsui}}\ and\ \bibinfo {author} {\bibfnamefont {H.}~\bibnamefont {Satz}},\
  }\bibfield  {journal} {\bibinfo  {journal} {Phys. Lett.}\ }\textbf {\bibinfo
  {volume} {B178}},\ \bibinfo {pages} {416} (\bibinfo {year}
  {1986})\BibitemShut {NoStop}%
%%CITATION = PHLTA,B178,416;%%
\bibitem [{\citenamefont {Abreu}\ \emph {et~al.}(2000)\citenamefont {Abreu}
  \emph {et~al.}}]{Abreu:2000ni}%
  \BibitemOpen
  \bibfield  {author} {\bibinfo {author} {\bibfnamefont {M.~C.}\ \bibnamefont
  {Abreu}} \emph {et~al.} (\bibinfo {collaboration} {NA50}),\ }\bibfield
  {journal} {\bibinfo  {journal} {Phys. Lett.}\ }\textbf {\bibinfo {volume}
  {B477}},\ \bibinfo {pages} {28} (\bibinfo {year} {2000})\BibitemShut
  {NoStop}%
%%CITATION = PHLTA,B477,28;%%
\bibitem [{\citenamefont {Arnaldi}\ \emph {et~al.}(2007)\citenamefont {Arnaldi}
  \emph {et~al.}}]{Arnaldi:2007zz}%
  \BibitemOpen
  \bibfield  {author} {\bibinfo {author} {\bibfnamefont {R.}~\bibnamefont
  {Arnaldi}} \emph {et~al.} (\bibinfo {collaboration} {NA60}),\ }\bibfield
  {journal} {\bibinfo  {journal} {Phys. Rev. Lett.}\ }\textbf {\bibinfo
  {volume} {99}},\ \bibinfo {pages} {132302} (\bibinfo {year}
  {2007})\BibitemShut {NoStop}%
%%CITATION = PRLTA,99,132302;%%
\bibitem [{\citenamefont {Chatrchyan}\ \emph {et~al.}(2011)\citenamefont
  {Chatrchyan} \emph {et~al.}}]{Chatrchyan:2011pe}%
  \BibitemOpen
  \bibfield  {author} {\bibinfo {author} {\bibfnamefont {S.}~\bibnamefont
  {Chatrchyan}} \emph {et~al.} (\bibinfo {collaboration} {CMS}),\ }\bibfield
  {journal} {\bibinfo  {journal} {Phys. Rev. Lett.}\ }\textbf {\bibinfo
  {volume} {107}},\ \bibinfo {pages} {052302} (\bibinfo {year} {2011}),\
  1105.4894\BibitemShut {NoStop}%
%%CITATION = ARXIV:1105.4894;%%
\bibitem [{\citenamefont {Bazavov}\ \emph {et~al.}(2016)\citenamefont
  {Bazavov}, \citenamefont {Petreczky},\ and\ \citenamefont
  {Weber}}]{Bazavov:2016qod}%
  \BibitemOpen
  \bibfield  {author} {\bibinfo {author} {\bibfnamefont {A.}~\bibnamefont
  {Bazavov}}, \bibinfo {author} {\bibfnamefont {P.}~\bibnamefont {Petreczky}},
  \ and\ \bibinfo {author} {\bibfnamefont {J.~H.}\ \bibnamefont {Weber}}\ }
  (\bibinfo {year} {2016}),\ 1601.08001\BibitemShut {NoStop}%
%%CITATION = ARXIV:1601.08001;%%
\bibitem [{\citenamefont {Borsanyi}\ \emph {et~al.}(2015)\citenamefont
  {Borsanyi}, \citenamefont {Fodor}, \citenamefont {Katz}, \citenamefont
  {Pasztor}, \citenamefont {Szabo},\ and\ \citenamefont
  {T{\"o}r{\"o}k}}]{Borsanyi:2015yka}%
  \BibitemOpen
  \bibfield  {author} {\bibinfo {author} {\bibfnamefont {S.}~\bibnamefont
  {Borsanyi}}, \bibinfo {author} {\bibfnamefont {Z.}~\bibnamefont {Fodor}},
  \bibinfo {author} {\bibfnamefont {S.~D.}\ \bibnamefont {Katz}}, \bibinfo
  {author} {\bibfnamefont {A.}~\bibnamefont {Pasztor}}, \bibinfo {author}
  {\bibfnamefont {K.~K.}\ \bibnamefont {Szabo}}, \ and\ \bibinfo {author}
  {\bibfnamefont {C.}~\bibnamefont {T{\"o}r{\"o}k}},\ }\bibfield  {journal}
  {\bibinfo  {journal} {JHEP}\ }\textbf {\bibinfo {volume} {04}},\ \bibinfo
  {pages} {138} (\bibinfo {year} {2015}),\ 1501.02173\BibitemShut {NoStop}%
%%CITATION = ARXIV:1501.02173;%%
\bibitem [{\citenamefont {Adler}\ \emph {et~al.}(2005)\citenamefont {Adler}
  \emph {et~al.}}]{Adler:2005ab}%
  \BibitemOpen
  \bibfield  {author} {\bibinfo {author} {\bibfnamefont {S.~S.}\ \bibnamefont
  {Adler}} \emph {et~al.} (\bibinfo {collaboration} {PHENIX}),\ }\bibfield
  {journal} {\bibinfo  {journal} {Phys. Rev.}\ }\textbf {\bibinfo {volume}
  {C72}},\ \bibinfo {pages} {024901} (\bibinfo {year} {2005}),\
  nucl-ex/0502009\BibitemShut {NoStop}%
%%CITATION = NUCL-EX/0502009;%%
\bibitem [{\citenamefont {Moore}\ and\ \citenamefont
  {Teaney}(2005)}]{Moore:2004tg}%
  \BibitemOpen
  \bibfield  {author} {\bibinfo {author} {\bibfnamefont {G.~D.}\ \bibnamefont
  {Moore}}\ and\ \bibinfo {author} {\bibfnamefont {D.}~\bibnamefont {Teaney}},\
  }\bibfield  {journal} {\bibinfo  {journal} {Phys. Rev.}\ }\textbf {\bibinfo
  {volume} {C71}},\ \bibinfo {pages} {064904} (\bibinfo {year} {2005}),\
  hep-ph/0412346\BibitemShut {NoStop}%
%%CITATION = HEP-PH/0412346;%%
\bibitem [{\citenamefont {van Hees}\ and\ \citenamefont
  {Rapp}(2005)}]{vanHees:2004gq}%
  \BibitemOpen
  \bibfield  {author} {\bibinfo {author} {\bibfnamefont {H.}~\bibnamefont {van
  Hees}}\ and\ \bibinfo {author} {\bibfnamefont {R.}~\bibnamefont {Rapp}},\
  }\bibfield  {journal} {\bibinfo  {journal} {Phys. Rev.}\ }\textbf {\bibinfo
  {volume} {C71}},\ \bibinfo {pages} {034907} (\bibinfo {year} {2005}),\
  nucl-th/0412015\BibitemShut {NoStop}%
%%CITATION = NUCL-TH/0412015;%%
\bibitem [{\citenamefont {van Hees}\ \emph {et~al.}(2006)\citenamefont {van
  Hees}, \citenamefont {Greco},\ and\ \citenamefont {Rapp}}]{vanHees:2005wb}%
  \BibitemOpen
  \bibfield  {author} {\bibinfo {author} {\bibfnamefont {H.}~\bibnamefont {van
  Hees}}, \bibinfo {author} {\bibfnamefont {V.}~\bibnamefont {Greco}}, \ and\
  \bibinfo {author} {\bibfnamefont {R.}~\bibnamefont {Rapp}},\ }\bibfield
  {journal} {\bibinfo  {journal} {Phys. Rev.}\ }\textbf {\bibinfo {volume}
  {C73}},\ \bibinfo {pages} {034913} (\bibinfo {year} {2006}),\
  nucl-th/0508055\BibitemShut {NoStop}%
%%CITATION = NUCL-TH/0508055;%%
\bibitem [{\citenamefont {Svetitsky}(1988)}]{Svetitsky:1987gq}%
  \BibitemOpen
  \bibfield  {author} {\bibinfo {author} {\bibfnamefont {B.}~\bibnamefont
  {Svetitsky}},\ }\bibfield  {journal} {\bibinfo  {journal} {Phys. Rev.}\
  }\textbf {\bibinfo {volume} {D37}},\ \bibinfo {pages} {2484} (\bibinfo {year}
  {1988})\BibitemShut {NoStop}%
%%CITATION = PHRVA,D37,2484;%%
\bibitem [{\citenamefont {Young}\ and\ \citenamefont
  {Shuryak}(2009)}]{Young:2008he}%
  \BibitemOpen
  \bibfield  {author} {\bibinfo {author} {\bibfnamefont {C.}~\bibnamefont
  {Young}}\ and\ \bibinfo {author} {\bibfnamefont {E.}~\bibnamefont
  {Shuryak}},\ }\bibfield  {journal} {\bibinfo  {journal} {Phys. Rev.}\
  }\textbf {\bibinfo {volume} {C79}},\ \bibinfo {pages} {034907} (\bibinfo
  {year} {2009}),\ 0803.2866\BibitemShut {NoStop}%
%%CITATION = ARXIV:0803.2866;%%
\bibitem [{\citenamefont {Young}\ and\ \citenamefont
  {Shuryak}(2010)}]{Young:2009tj}%
  \BibitemOpen
  \bibfield  {author} {\bibinfo {author} {\bibfnamefont {C.}~\bibnamefont
  {Young}}\ and\ \bibinfo {author} {\bibfnamefont {E.}~\bibnamefont
  {Shuryak}},\ }\bibfield  {journal} {\bibinfo  {journal} {Phys. Rev.}\
  }\textbf {\bibinfo {volume} {C81}},\ \bibinfo {pages} {034905} (\bibinfo
  {year} {2010}),\ 0911.3080\BibitemShut {NoStop}%
%%CITATION = ARXIV:0911.3080;%%
\bibitem [{\citenamefont {Young}\ \emph {et~al.}(2012)\citenamefont {Young},
  \citenamefont {Schenke}, \citenamefont {Jeon},\ and\ \citenamefont
  {Gale}}]{Young:2011ug}%
  \BibitemOpen
  \bibfield  {author} {\bibinfo {author} {\bibfnamefont {C.}~\bibnamefont
  {Young}}, \bibinfo {author} {\bibfnamefont {B.}~\bibnamefont {Schenke}},
  \bibinfo {author} {\bibfnamefont {S.}~\bibnamefont {Jeon}}, \ and\ \bibinfo
  {author} {\bibfnamefont {C.}~\bibnamefont {Gale}},\ }\bibfield  {journal}
  {\bibinfo  {journal} {Phys. Rev.}\ }\textbf {\bibinfo {volume} {C86}},\
  \bibinfo {pages} {034905} (\bibinfo {year} {2012}),\ 1111.0647\BibitemShut
  {NoStop}%
%%CITATION = ARXIV:1111.0647;%%
\bibitem [{\citenamefont {Sjostrand}\ \emph {et~al.}(2008)\citenamefont
  {Sjostrand}, \citenamefont {Mrenna},\ and\ \citenamefont
  {Skands}}]{Sjostrand:2007gs}%
  \BibitemOpen
  \bibfield  {author} {\bibinfo {author} {\bibfnamefont {T.}~\bibnamefont
  {Sjostrand}}, \bibinfo {author} {\bibfnamefont {S.}~\bibnamefont {Mrenna}}, \
  and\ \bibinfo {author} {\bibfnamefont {P.~Z.}\ \bibnamefont {Skands}},\
  }\bibfield  {journal} {\bibinfo  {journal} {Comput. Phys. Commun.}\ }\textbf
  {\bibinfo {volume} {178}},\ \bibinfo {pages} {852} (\bibinfo {year} {2008}),\
  0710.3820\BibitemShut {NoStop}%
%%CITATION = ARXIV:0710.3820;%%
\bibitem [{\citenamefont {Fritzsch}\ and\ \citenamefont
  {Streng}(1978)}]{Fritzsch:1978kn}%
  \BibitemOpen
  \bibfield  {author} {\bibinfo {author} {\bibfnamefont {H.}~\bibnamefont
  {Fritzsch}}\ and\ \bibinfo {author} {\bibfnamefont {K.~H.}\ \bibnamefont
  {Streng}},\ }\bibfield  {journal} {\bibinfo  {journal} {Phys. Lett.}\
  }\textbf {\bibinfo {volume} {B78}},\ \bibinfo {pages} {447} (\bibinfo {year}
  {1978})\BibitemShut {NoStop}%
%%CITATION = PHLTA,B78,447;%%
\bibitem [{\citenamefont {Amundson}\ \emph {et~al.}(1997)\citenamefont
  {Amundson}, \citenamefont {Eboli}, \citenamefont {Gregores},\ and\
  \citenamefont {Halzen}}]{Amundson:1996qr}%
  \BibitemOpen
  \bibfield  {author} {\bibinfo {author} {\bibfnamefont {J.~F.}\ \bibnamefont
  {Amundson}}, \bibinfo {author} {\bibfnamefont {O.~J.~P.}\ \bibnamefont
  {Eboli}}, \bibinfo {author} {\bibfnamefont {E.~M.}\ \bibnamefont {Gregores}},
  \ and\ \bibinfo {author} {\bibfnamefont {F.}~\bibnamefont {Halzen}},\
  }\bibfield  {journal} {\bibinfo  {journal} {Phys. Lett.}\ }\textbf {\bibinfo
  {volume} {B390}},\ \bibinfo {pages} {323} (\bibinfo {year} {1997}),\
  hep-ph/9605295\BibitemShut {NoStop}%
%%CITATION = HEP-PH/9605295;%%
\bibitem [{\citenamefont {Casalderrey-Solana}\ and\ \citenamefont
  {Teaney}(2006)}]{CasalderreySolana:2006rq}%
  \BibitemOpen
  \bibfield  {author} {\bibinfo {author} {\bibfnamefont {J.}~\bibnamefont
  {Casalderrey-Solana}}\ and\ \bibinfo {author} {\bibfnamefont
  {D.}~\bibnamefont {Teaney}},\ }\bibfield  {journal} {\bibinfo  {journal}
  {Phys. Rev.}\ }\textbf {\bibinfo {volume} {D74}},\ \bibinfo {pages} {085012}
  (\bibinfo {year} {2006}),\ hep-ph/0605199\BibitemShut {NoStop}%
%%CITATION = HEP-PH/0605199;%%
\bibitem [{\citenamefont {Akamatsu}(2015)}]{Akamatsu:2014qsa}%
  \BibitemOpen
  \bibfield  {author} {\bibinfo {author} {\bibfnamefont {Y.}~\bibnamefont
  {Akamatsu}},\ }\bibfield  {journal} {\bibinfo  {journal} {Phys. Rev.}\
  }\textbf {\bibinfo {volume} {D91}},\ \bibinfo {pages} {056002} (\bibinfo
  {year} {2015}),\ 1403.5783\BibitemShut {NoStop}%
%%CITATION = ARXIV:1403.5783;%%
\bibitem [{\citenamefont {Petrov}(2006)}]{Petrov:2006pf}%
  \BibitemOpen
  \bibfield  {author} {\bibinfo {author} {\bibfnamefont {K.}~\bibnamefont
  {Petrov}} (\bibinfo {collaboration} {RBC-Bielefeld}),\ }\bibfield
  {booktitle} {\emph {\bibinfo {booktitle} {{Proceedings, 24th International
  Symposium on Lattice Field Theory (Lattice 2006)}}},\ }\bibfield  {journal}
  {\bibinfo  {journal} {PoS}\ }\textbf {\bibinfo {volume} {LAT2006}},\ \bibinfo
  {pages} {144} (\bibinfo {year} {2006}),\ hep-lat/0610041\BibitemShut
  {NoStop}%
%%CITATION = HEP-LAT/0610041;%%
\bibitem [{\citenamefont {Bazavov}\ \emph {et~al.}(2014)\citenamefont
  {Bazavov}, \citenamefont {Burnier},\ and\ \citenamefont
  {Petreczky}}]{Bazavov:2014kva}%
  \BibitemOpen
  \bibfield  {author} {\bibinfo {author} {\bibfnamefont {A.}~\bibnamefont
  {Bazavov}}, \bibinfo {author} {\bibfnamefont {Y.}~\bibnamefont {Burnier}}, \
  and\ \bibinfo {author} {\bibfnamefont {P.}~\bibnamefont {Petreczky}},\
  }\bibfield  {booktitle} {\emph {\bibinfo {booktitle} {{Proceedings, 6th
  International Conference on Hard and Electromagnetic Probes of High-Energy
  Nuclear Collisions (Hard Probes 2013)}}},\ }\bibfield  {journal} {\bibinfo
  {journal} {Nucl. Phys.}\ }\textbf {\bibinfo {volume} {A932}},\ \bibinfo
  {pages} {117} (\bibinfo {year} {2014}),\ 1404.4267\BibitemShut {NoStop}%
%%CITATION = ARXIV:1404.4267;%%
\bibitem [{\citenamefont {Burnier}\ \emph {et~al.}(2015)\citenamefont
  {Burnier}, \citenamefont {Kaczmarek},\ and\ \citenamefont
  {Rothkopf}}]{Burnier:2014ssa}%
  \BibitemOpen
  \bibfield  {author} {\bibinfo {author} {\bibfnamefont {Y.}~\bibnamefont
  {Burnier}}, \bibinfo {author} {\bibfnamefont {O.}~\bibnamefont {Kaczmarek}},
  \ and\ \bibinfo {author} {\bibfnamefont {A.}~\bibnamefont {Rothkopf}},\
  }\bibfield  {journal} {\bibinfo  {journal} {Phys. Rev. Lett.}\ }\textbf
  {\bibinfo {volume} {114}},\ \bibinfo {pages} {082001} (\bibinfo {year}
  {2015}),\ 1410.2546\BibitemShut {NoStop}%
%%CITATION = ARXIV:1410.2546;%%
\bibitem [{\citenamefont {Mocsy}\ and\ \citenamefont
  {Petreczky}(2007)}]{Mocsy:2007jz}%
  \BibitemOpen
  \bibfield  {author} {\bibinfo {author} {\bibfnamefont {A.}~\bibnamefont
  {Mocsy}}\ and\ \bibinfo {author} {\bibfnamefont {P.}~\bibnamefont
  {Petreczky}},\ }\bibfield  {journal} {\bibinfo  {journal} {Phys. Rev. Lett.}\
  }\textbf {\bibinfo {volume} {99}},\ \bibinfo {pages} {211602} (\bibinfo
  {year} {2007}),\ 0706.2183\BibitemShut {NoStop}%
%%CITATION = ARXIV:0706.2183;%%
\bibitem [{\citenamefont {Bazavov}\ and\ \citenamefont
  {Petreczky}(2013)}]{Bazavov:2012fk}%
  \BibitemOpen
  \bibfield  {author} {\bibinfo {author} {\bibfnamefont {A.}~\bibnamefont
  {Bazavov}}\ and\ \bibinfo {author} {\bibfnamefont {P.}~\bibnamefont
  {Petreczky}},\ }\bibfield  {booktitle} {\emph {\bibinfo {booktitle}
  {{Proceedings, Extreme QCD 2012 (XQCD12)}}},\ }\bibfield  {journal} {\bibinfo
   {journal} {J. Phys. Conf. Ser.}\ }\textbf {\bibinfo {volume} {432}},\
  \bibinfo {pages} {012003} (\bibinfo {year} {2013}),\ 1211.5638\BibitemShut
  {NoStop}%
%%CITATION = ARXIV:1211.5638;%%
\bibitem [{\citenamefont {Brambilla}\ \emph {et~al.}(2008)\citenamefont
  {Brambilla}, \citenamefont {Ghiglieri}, \citenamefont {Vairo},\ and\
  \citenamefont {Petreczky}}]{Brambilla:2008cx}%
  \BibitemOpen
  \bibfield  {author} {\bibinfo {author} {\bibfnamefont {N.}~\bibnamefont
  {Brambilla}}, \bibinfo {author} {\bibfnamefont {J.}~\bibnamefont
  {Ghiglieri}}, \bibinfo {author} {\bibfnamefont {A.}~\bibnamefont {Vairo}}, \
  and\ \bibinfo {author} {\bibfnamefont {P.}~\bibnamefont {Petreczky}},\
  }\bibfield  {journal} {\bibinfo  {journal} {Phys. Rev.}\ }\textbf {\bibinfo
  {volume} {D78}},\ \bibinfo {pages} {014017} (\bibinfo {year} {2008}),\
  0804.0993\BibitemShut {NoStop}%
%%CITATION = ARXIV:0804.0993;%%
\bibitem [{\citenamefont {Brambilla}\ \emph {et~al.}(2010)\citenamefont
  {Brambilla}, \citenamefont {Escobedo}, \citenamefont {Ghiglieri},
  \citenamefont {Soto},\ and\ \citenamefont {Vairo}}]{Brambilla:2010vq}%
  \BibitemOpen
  \bibfield  {author} {\bibinfo {author} {\bibfnamefont {N.}~\bibnamefont
  {Brambilla}}, \bibinfo {author} {\bibfnamefont {M.~A.}\ \bibnamefont
  {Escobedo}}, \bibinfo {author} {\bibfnamefont {J.}~\bibnamefont {Ghiglieri}},
  \bibinfo {author} {\bibfnamefont {J.}~\bibnamefont {Soto}}, \ and\ \bibinfo
  {author} {\bibfnamefont {A.}~\bibnamefont {Vairo}},\ }\bibfield  {journal}
  {\bibinfo  {journal} {JHEP}\ }\textbf {\bibinfo {volume} {09}},\ \bibinfo
  {pages} {038} (\bibinfo {year} {2010}),\ 1007.4156\BibitemShut {NoStop}%
%%CITATION = ARXIV:1007.4156;%%
\bibitem [{\citenamefont {Brambilla}\ \emph {et~al.}(2011)\citenamefont
  {Brambilla}, \citenamefont {Escobedo}, \citenamefont {Ghiglieri},\ and\
  \citenamefont {Vairo}}]{Brambilla:2011sg}%
  \BibitemOpen
  \bibfield  {author} {\bibinfo {author} {\bibfnamefont {N.}~\bibnamefont
  {Brambilla}}, \bibinfo {author} {\bibfnamefont {M.~A.}\ \bibnamefont
  {Escobedo}}, \bibinfo {author} {\bibfnamefont {J.}~\bibnamefont {Ghiglieri}},
  \ and\ \bibinfo {author} {\bibfnamefont {A.}~\bibnamefont {Vairo}},\
  }\bibfield  {journal} {\bibinfo  {journal} {JHEP}\ }\textbf {\bibinfo
  {volume} {12}},\ \bibinfo {pages} {116} (\bibinfo {year} {2011}),\
  1109.5826\BibitemShut {NoStop}%
%%CITATION = ARXIV:1109.5826;%%
\bibitem [{\citenamefont {Brambilla}\ \emph {et~al.}(2013)\citenamefont
  {Brambilla}, \citenamefont {Escobedo}, \citenamefont {Ghiglieri},\ and\
  \citenamefont {Vairo}}]{Brambilla:2013dpa}%
  \BibitemOpen
  \bibfield  {author} {\bibinfo {author} {\bibfnamefont {N.}~\bibnamefont
  {Brambilla}}, \bibinfo {author} {\bibfnamefont {M.~A.}\ \bibnamefont
  {Escobedo}}, \bibinfo {author} {\bibfnamefont {J.}~\bibnamefont {Ghiglieri}},
  \ and\ \bibinfo {author} {\bibfnamefont {A.}~\bibnamefont {Vairo}},\
  }\bibfield  {journal} {\bibinfo  {journal} {JHEP}\ }\textbf {\bibinfo
  {volume} {05}},\ \bibinfo {pages} {130} (\bibinfo {year} {2013}),\
  1303.6097\BibitemShut {NoStop}%
%%CITATION = ARXIV:1303.6097;%%
\bibitem [{\citenamefont {Young}\ and\ \citenamefont
  {Dusling}(2013)}]{Young:2010jq}%
  \BibitemOpen
  \bibfield  {author} {\bibinfo {author} {\bibfnamefont {C.}~\bibnamefont
  {Young}}\ and\ \bibinfo {author} {\bibfnamefont {K.}~\bibnamefont
  {Dusling}},\ }\bibfield  {journal} {\bibinfo  {journal} {Phys. Rev.}\
  }\textbf {\bibinfo {volume} {C87}},\ \bibinfo {pages} {065206} (\bibinfo
  {year} {2013}),\ 1001.0935\BibitemShut {NoStop}%
%%CITATION = ARXIV:1001.0935;%%
\bibitem [{\citenamefont {Buyukdag}\ and\ \citenamefont
  {Young}(2015)}]{Buyukdag:2015sba}%
  \BibitemOpen
  \bibfield  {author} {\bibinfo {author} {\bibfnamefont {Y.}~\bibnamefont
  {Buyukdag}}\ and\ \bibinfo {author} {\bibfnamefont {C.}~\bibnamefont
  {Young}},\ }\bibfield  {journal} {\bibinfo  {journal} {Phys. Rev.}\ }\textbf
  {\bibinfo {volume} {C91}},\ \bibinfo {pages} {045204} (\bibinfo {year}
  {2015}),\ 1504.00343\BibitemShut {NoStop}%
%%CITATION = ARXIV:1504.00343;%%
\bibitem [{\citenamefont {Akamatsu}\ and\ \citenamefont
  {Rothkopf}(2012)}]{Akamatsu:2011se}%
  \BibitemOpen
  \bibfield  {author} {\bibinfo {author} {\bibfnamefont {Y.}~\bibnamefont
  {Akamatsu}}\ and\ \bibinfo {author} {\bibfnamefont {A.}~\bibnamefont
  {Rothkopf}},\ }\bibfield  {journal} {\bibinfo  {journal} {Phys. Rev.}\
  }\textbf {\bibinfo {volume} {D85}},\ \bibinfo {pages} {105011} (\bibinfo
  {year} {2012}),\ 1110.1203\BibitemShut {NoStop}%
%%CITATION = ARXIV:1110.1203;%%
\bibitem [{\citenamefont {Rothkopf}(2013)}]{Rothkopf:2013ria}%
  \BibitemOpen
  \bibfield  {author} {\bibinfo {author} {\bibfnamefont {A.}~\bibnamefont
  {Rothkopf}},\ }\bibfield  {journal} {\bibinfo  {journal} {Mod. Phys. Lett.}\
  }\textbf {\bibinfo {volume} {A28}},\ \bibinfo {pages} {1330005} (\bibinfo
  {year} {2013}),\ 1302.6195\BibitemShut {NoStop}%
%%CITATION = ARXIV:1302.6195;%%
\bibitem [{\citenamefont {Rothkopf}(2014)}]{Rothkopf:2013kya}%
  \BibitemOpen
  \bibfield  {author} {\bibinfo {author} {\bibfnamefont {A.}~\bibnamefont
  {Rothkopf}},\ }\bibfield  {journal} {\bibinfo  {journal} {JHEP}\ }\textbf
  {\bibinfo {volume} {04}},\ \bibinfo {pages} {085} (\bibinfo {year} {2014}),\
  1312.3246\BibitemShut {NoStop}%
%%CITATION = ARXIV:1312.3246;%%
\bibitem [{\citenamefont {Blaizot}\ \emph {et~al.}(2016)\citenamefont
  {Blaizot}, \citenamefont {De~Boni}, \citenamefont {Faccioli},\ and\
  \citenamefont {Garberoglio}}]{Blaizot:2015hya}%
  \BibitemOpen
  \bibfield  {author} {\bibinfo {author} {\bibfnamefont {J.-P.}\ \bibnamefont
  {Blaizot}}, \bibinfo {author} {\bibfnamefont {D.}~\bibnamefont {De~Boni}},
  \bibinfo {author} {\bibfnamefont {P.}~\bibnamefont {Faccioli}}, \ and\
  \bibinfo {author} {\bibfnamefont {G.}~\bibnamefont {Garberoglio}},\
  }\bibfield  {journal} {\bibinfo  {journal} {Nucl. Phys.}\ }\textbf {\bibinfo
  {volume} {A946}},\ \bibinfo {pages} {49} (\bibinfo {year} {2016}),\
  1503.03857\BibitemShut {NoStop}%
%%CITATION = ARXIV:1503.03857;%%
\bibitem [{\citenamefont {Gossiaux}\ and\ \citenamefont
  {Katz}(2016)}]{Gossiaux:2016htk}%
  \BibitemOpen
  \bibfield  {author} {\bibinfo {author} {\bibfnamefont {P.~B.}\ \bibnamefont
  {Gossiaux}}\ and\ \bibinfo {author} {\bibfnamefont {R.}~\bibnamefont
  {Katz}},\ }in\ \href
  {http://inspirehep.net/record/1414193/files/arXiv:1601.01443.pdf} {\emph
  {\bibinfo {booktitle} {{25th International Conference on Ultra-Relativistic
  Nucleus-Nucleus Collisions (Quark Matter 2015) Kobe, Japan, September
  27-October 3, 2015}}}}\ (\bibinfo {year} {2016})\ 1601.01443\BibitemShut
  {NoStop}%
%%CITATION = ARXIV:1601.01443;%%
\bibitem [{\citenamefont {Kim}\ \emph {et~al.}(2015)\citenamefont {Kim},
  \citenamefont {Petreczky},\ and\ \citenamefont {Rothkopf}}]{Kim:2014iga}%
  \BibitemOpen
  \bibfield  {author} {\bibinfo {author} {\bibfnamefont {S.}~\bibnamefont
  {Kim}}, \bibinfo {author} {\bibfnamefont {P.}~\bibnamefont {Petreczky}}, \
  and\ \bibinfo {author} {\bibfnamefont {A.}~\bibnamefont {Rothkopf}},\
  }\bibfield  {journal} {\bibinfo  {journal} {Phys. Rev.}\ }\textbf {\bibinfo
  {volume} {D91}},\ \bibinfo {pages} {054511} (\bibinfo {year} {2015}),\
  1409.3630\BibitemShut {NoStop}%
%%CITATION = ARXIV:1409.3630;%%
\bibitem [{\citenamefont {Mukherjee}\ \emph {et~al.}(2016)\citenamefont
  {Mukherjee}, \citenamefont {Petreczky},\ and\ \citenamefont
  {Sharma}}]{Mukherjee:2015mxc}%
  \BibitemOpen
  \bibfield  {author} {\bibinfo {author} {\bibfnamefont {S.}~\bibnamefont
  {Mukherjee}}, \bibinfo {author} {\bibfnamefont {P.}~\bibnamefont
  {Petreczky}}, \ and\ \bibinfo {author} {\bibfnamefont {S.}~\bibnamefont
  {Sharma}},\ }\bibfield  {journal} {\bibinfo  {journal} {Phys. Rev.}\ }\textbf
  {\bibinfo {volume} {D93}},\ \bibinfo {pages} {014502} (\bibinfo {year}
  {2016}),\ 1509.08887\BibitemShut {NoStop}%
%%CITATION = ARXIV:1509.08887;%%
\end{thebibliography}%
